\documentstyle[aps,pre,eqsecnum]{revtex}
\begin{document}

\title{A Path Integral Approach to Effective Non-linear Medium}

\author{Marc Barth\'el\'emy$^{(1)}$ and Henri Orland$^{(2)}$}
\address{$^{(1)}$Commissariat \`a l'Energie Atomique, Service de Physique de la Mati\`ere
Condens\'ee\\BP12, 91680 Bruy\`eres-Le-Chatel, France}
\address{$^{(2)}$Commissariat \`a l'Energie Atomique, Service de Physique Th\'eorique\\ 91191 Gif-Sur-Yvette Cedex, France}

\date{\today}

\maketitle

\def\be{\begin{equation}}
\def\ee{\end{equation}}
\def\vr{\bf r}
\def\ddx{\text{d}^d{x}}
\def\ddu{\text{d}^d{\bf u}}
\def\vk{\bf k}
\def\tln{\text{ln}}
\newcommand{\vrp}{\bf r'}
\newcommand{\gti}{\tilde g}
 \newcommand{\ddk}{\frac{d^d k}{(2 \pi)^d}}

\vskip 1cm

\begin{abstract}
In this article, we propose a new method to compute the effective properties of non-linear disordered media. We use the fact that the effective constants can be defined through the minimum of an energy functional. We express this minimum in terms of a path integral allowing us to use many-body techniques. We obtain the perturbation expansion of the effective constants to second order in disorder, for any kind of non-linearity. We apply our method to the case of strong non-linearities (i.e. $D=\varepsilon(E^2)^{\kappa/2}E$, where $\varepsilon$ is fluctuating from point to point), and to the case of weak non-linearity (i.e. $D=\varepsilon E+\chi(E^2) E$ where $\varepsilon$ and $\chi$ fluctuate from point to point).
Our results are in agreement with previous ones, and could be easily extended to other types of non-linear problems in disordered systems.
\end{abstract}

\vskip 1cm\noindent\mbox{PACS: 72.20Ht, 77.84Lf} \hfill
\vskip 1cm\noindent\mbox{Submitted for publication to: European Physical Journal B}
\hfill


\section{Introduction}
\label{sec: intro}

The study of the properties of linear heterogeneous media (such as composites, suspensions) has been the subject of an intense activity for already fifty years (see the reviews \cite{landauer} and \cite{BergmanSSP}). More recently there has been a great interest in non-linear media \cite{BergmanSSP}. The non-linearities appear in strong applied fields. In that case, one must expand the local generalized susceptibility in powers of the applied field. If the system is invariant under inversion symmetries \cite{landau}, one can write a relation between the displacement field $D$ and the electric field $E$ in the form
\be
D=\varepsilon E+\chi(E^2)E
\ee
or in the metallic case
\be
j=\sigma E+\chi(E^2)E
\ee
In some cases, the material does not possess a linear regime, and the local equation becomes
\be
D=\chi(E^2)^{\kappa/2}E
\ee
with $\kappa\ge -1$ (and a similar equation in the metallic case). In the former case, we can speak of `weak' non-linearity (WNL) and in the latter one of `strong' non-linearity (SNL). In the preceding equations,  the quantities $\varepsilon$ and $\chi$ are position-dependent and fluctuate randomly from point to point.\\
It has been shown \cite{stroudhui88} that the WNL case is connected to conductivity fluctuations, the so-called flicker noise \cite{BergmanSSP}. It has also been shown \cite{stroudhui88}, that in the limit of weak non-linearity (i.e. $\chi E^2\ll\varepsilon$), the effective non-linear susceptibility $\chi^*$ is related to the fourth moment of the local electric field computed in the {\it linear} problem (i.e. $\chi=0$). This shows in particular that the effective non-linear susceptibility is very sensitive to the micro-structure and that one should be careful in building effective medium approximations \cite{bergman89}.
This result allowed Stroud and Hui to compute the effective susceptibility to first order in impurity concentration. In addition, Zeng et al \cite{zeng88} used this relation between the fourth moment of the electric field and the effective non-linear susceptibility in order to propose an effective medium approximation (EMA). This approximation relies essentially on the `factorization approximation' (which assumes that the local electric field is Gaussian distributed, a questionable hypothesis, especially near the percolation threshold). The critical behavior (near the percolation threshold) was investigated in details by Levy and Bergman \cite{levy} and numerical simulations were done in \cite{zhang94}, \cite{hui95}. In the case where the non-linear term behaves as $\chi(E^2)^{\kappa/2}E$, Hui and Chung gave a generalization of the effective medium approximation \cite{huichung95}.\\
In the SNL case, the first important result is due to Blumenfeld and Bergman \cite{blumen1}, \cite{blumen2} who derived the second order perturbation theory of the effective non-linear susceptibility for three-dimensional composites. This result was extended to any dimension $d$ \cite{lee}. Ponte-Casta\~neda \cite{ponte92} proposed a variational approximation in order to establish optimal bounds. More recently, EMAs (and numerical simulations) were developed in this strongly non-linear case \cite{wan96}, \cite{gao96}, \cite{sali97}, \cite{levy2}.\\
For both cases, the EMA  reproduces at least qualitatively the numerical results, but there is still a lack of clarity in the approximations used. The method we propose here could serve as a starting point for alternative derivations of EMA's, and possesses the advantage of not being limited to a special kind of non-linearities. Moreover, we believe that this method could be applied to the celebrated random fuse network problem (see for example the review \cite{roux}). The essential point in our method is that one can define the effective constant through the minimum (with some constraints) of the stored (or dissipated) energy. We then express this (constrained) minimum in terms of a path integral, allowing us to apply usual methods from many-body theory. We derive the second-order perturbation theory for any types of non-linearities (leaving more refined approximations for the future \cite{boprepa}). We illustrate our results on both the SNL and WNL cases.\\
In the next part, we present the basic equations and derive the second-order perturbation theory. In (\ref{SNL}) and (\ref{WNL}), we apply our method to the SNL and WNL cases.


\section{Calculations}
\label{sec: calc}

\subsection{Basic equations and second-order perturbation theory}

The effective constants are usually defined through the following relations. In the WNL case, we have
\be
\langle D\rangle=\varepsilon^*E_0+\chi^*(E_0^2)E_0
\ee
where $E_0$ is the average field, while in the SNL case
\be
\langle D\rangle=\chi^*(E_0^2)^{\kappa/2}E_0
\ee
We first note that there is an alternative definition of the effective constants (see \cite{ponte92} and references therein). Let us denote by $\sigma_i,\ i=1,...,d$ the generic zero-divergence d-dimensional field. In the dielectric case $\sigma$ is the displacement field $\sigma_i=D_i$ whereas in the conducting case it is the current $\sigma_i=j_i$. The stored energy $W$ (or dissipated energy in the conducting case) in the system is given in terms of the energy density $w$ by

\be
W=\int {\ddx} w({\sigma}(x))
\ee
Hereafter, we take the volume of the sample equal to one. For heterogeneous materials, the energy density depends on random quantities fluctuating from point to point. We will specialize our discussion to binary disorder for which the parameters can take only two values (our method can easily be generalized to other types of disorder). We thus assume that the energy density, at each point, is distributed according to the following distribution probability
\be
P(w=w(\sigma (x)))=p\delta (w-w_1(\sigma (x)))+q\delta (w-w_2(\sigma (x)))
\ee
The alternative of solving Maxwell's equations is to minimize the total energy $W$ subjected to the two constraints $\nabla\cdot\sigma=0$ and $\overline\sigma_i=\Upsilon_i$ \cite{ponte92} (the bar denotes the spatial average). As is the case for the free energy in disordered systems, one expects this minimum to be self-averaging, allowing us to compute its average over disorder

\be
\langle\text{min}_
{\overline{\sigma}=\Upsilon
\atop{\nabla\cdot\sigma=0}}
\int\ddx w(\sigma)\rangle=W^*(\Upsilon)
\ee
where the brackets $\langle\cdot\rangle$ denotes the average over disorder and  $W^*$ denotes the stored energy in a homogeneous medium characterized by effective constants. The problem thus reduces to the calculation of the average of the minimum of a functional of the field subjected to the constraints of zero divergence and fixed mean value. The main point is to rewrite this constrained minimum as a path integral
\begin{equation}
\text{min}_{{\overline\sigma=\Upsilon\atop{\nabla}\cdot\sigma=0}}\int\ddx w(\sigma(x))=\text{lim}_{\beta\rightarrow\infty}-\frac{1}{\beta}\text{ln}\int{\cal D}{\sigma}\delta({\nabla}\cdot\sigma)\delta(\overline{\sigma}-\Upsilon)e^{-\beta\int w(\sigma(x))}
\end{equation}
The minimum can thus be interpreted as the ground state energy associated to the partition function $Z$ given by
\be
\label{ZZ}
Z=\int{\cal D}{\sigma}\delta({\nabla}\cdot\sigma)\delta(\overline{\sigma}-\Upsilon)e^{-\beta\int w(\sigma(x))}
\end{equation}
We have to compute the average of the logarithm of (\ref{ZZ}). 
In order to compute this quantity, we introduce replicas \cite{repliques} and use $\text{ln}Z=\text{lim}_{n\rightarrow 0}\frac{Z^n-1}{n}$. We thus have
\be
W^*=\text{lim}_{\beta\rightarrow\infty}\text{lim}_{n\rightarrow 0}-\frac{1}{\beta}\frac{\langle Z^n\rangle-1}{n}
\ee
Note that the order of the limits is important. The replica method relies on the fact that one can easily compute  $\langle Z^n\rangle$ for $n$ integer, and then take the limit $n\rightarrow 0$. The main quantity to study here is thus
\be
\langle Z^n\rangle=\int\prod_{\alpha=1}^{n}{\cal D}{\sigma}^{\alpha}
\delta({\nabla}\cdot\sigma^{\alpha})
\delta(\overline{\sigma}^{\alpha}-\Upsilon)
\langle e^{-\beta\int\sum_{\alpha=1}^{n} w(\sigma(x)^{\alpha})}\rangle
\end{equation}
which for a binary disorder reads
\be
\label{znmoyen}
\langle Z^n\rangle=\int\prod_{\alpha=1}^{n}{\cal D}\sigma^{\alpha}
\delta({\nabla}\cdot\sigma^{\alpha})
\delta(\overline{\sigma}^{\alpha}-\Upsilon)
\text{exp}\left[
\int\ddx\text{ln}
(pe^{-\beta\sum_{\alpha=1}^{n} w_1(\sigma^{\alpha}(x))}
+qe^{-\beta\sum_{\alpha=1}^{n} w_2(\sigma^{\alpha}(x))})
\right]
\end{equation}
Note that in this expression, a cut-off should appear; it will only play a role in eq. (\ref{wfinale}), and we will not write it in the intermediate expressions. We can now expand (\ref{znmoyen}) to second order in disorder. The fields $\sigma^{\alpha}(x)$ fluctuate around their mean value $\Upsilon$. To zeroth order, we take $\sigma^{\alpha}(x)=\Upsilon$ in equation (\ref{znmoyen}), leading to
\be
W^*(\Upsilon)=\langle w(\Upsilon)\rangle
\ee
In order to set up a perturbation theory to second order, we write  $\sigma^{\alpha}(x)=\Upsilon+{\epsilon}^{\alpha}$ and expand the exponential in (\ref{znmoyen})in powers of $\epsilon$. We obtain (repeated indices are summed)

\be
\label{pathintper}
\langle Z^n\rangle\simeq
e^{\text{ln}\langle e^{-\beta n w(\Upsilon)}\rangle}
\int\prod_{\alpha=1}^{n}{\cal D}\epsilon^{\alpha}_i
\delta({\nabla}\cdot{\epsilon}^{\alpha})
\delta(\overline{\epsilon}^{\alpha})
e^{-\beta\int\ddx
\epsilon_i^{\alpha}M_{ij}^{\alpha\gamma}\epsilon_j^{\gamma}
}
\end{equation}
where the tensor $M$ is given by
\be
M_{ij}^{\alpha\gamma}=\delta_{\alpha\gamma}[w''_{ij}]-\beta([w'_iw'_j]-[w'_i][w'_j])
\ee
where $w'_i$ denotes $\frac{d w}{d\sigma_i}$ at $\sigma=\Upsilon$ (and $w''_{ij}=\frac{d^2w}{d\sigma_i d\sigma_j}$ at $\sigma=\Upsilon$). The brackets  $[A]$ denote
\be
[A]=\frac
{\langle Ae^{-\beta n w(\Upsilon)}\rangle}
{\langle e^{-\beta n w(\Upsilon)}\rangle}
\ee
In the following, we will only need the zeroth order in $n$ of these quantities $[A]$, which is $[A]=\langle A\rangle +{\cal O}(n)$.\\
At this stage, we have to check that the tensor $M$ is positive definite. Such is the case when the term with the $\beta$ factor is sufficiently small. This term is related to the variance of the disorder, and this requirement coincides with the definition of the low contrast limit. We can perform the path integral (\ref{pathintper}) using the Fourier representation (see appendix A for details), and we obtain

\be
\label{perturb}
\langle Z^n\rangle\simeq
e^{\text{ln}\langle e^{-\beta n w(\Upsilon)}\rangle}
e^{-\frac{1}{2}\sum_{k\neq 0}\text{trln}M}
e^{-\frac{1}{2}\sum_{k\neq 0}\text{trln}(kM^{-1}k)}
\ee
(where inessential factors were omitted). The first trace is over both space indices $(i,j)$ and replica indices $(\alpha,\gamma)$ and the second over replica indices only.\\
Without loss of generality, we can assume that the energy density is a function of $\sigma^2$: $w=w(\sigma^2)$. In this case, the tensor $M$ can be written as
\be
\label{matrixM}
M_{ij}^{\alpha\gamma}=\delta_{\alpha\gamma}(\delta_{ij}A-B\Upsilon_i\Upsilon_j)-\beta C\Upsilon_i\Upsilon_j
\ee
Let us note that, due to spatial isotropy, the only tensors that can appear are $\delta_{ij}$ and $\Upsilon_i\Upsilon_j$.\\
One can then compute the determinants in expression (\ref{perturb}) and one obtains (for $n\rightarrow 0$ and to the first order in the variance of disorder)
\be
\frac{1}{n}\text{trln}M\simeq (d-1)\tln(A)+\tln(A-B\Upsilon^2)-\frac{\beta C\Upsilon^2}{A-B\Upsilon^2}
\ee
and
\be
\frac{1}{n}\text{trln}(kM^{-1}k)\simeq \tln(\frac{k^2}{A}-\frac{B(k\cdot\Upsilon)^2}{A(A-B\Upsilon^2})+\frac
{\beta AC\Upsilon^2}
{(A-B\Upsilon^2)((A-B\Upsilon^2)k^2-B(k\cdot\Upsilon)^2)}
\ee
The effective energy is given by the dominant term in $\beta$

\be
\label{wfinale}
W^*(\Upsilon)=\langle w(\Upsilon)\rangle-\frac{1}{2}\frac{C\Upsilon^2}{A-B\Upsilon^2}\left[1-\frac{A}{A-B\Upsilon^2}\frac{1}{(2\Lambda)^d}\sum_{k\ne 0}\frac{(k\cdot\hat{\Upsilon})^2}{k^2+U(k\cdot\hat{\Upsilon})^2}\right]
\ee
where $U=\frac{B\Upsilon^2}{A-B\Upsilon^2}$ and we have explicitly written the cut-off $\Lambda$ in Fourier space.\\
The sum over $k$ can be transformed into an integral (expressed in d-dimensional polar coordinates)

\begin{mathletters}
\begin{eqnarray}
\frac{1}{(2\Lambda)^d}\sum_{k\ne 0}f(\theta)
&=&\frac{d}{S_d\Lambda^d}\int_{0}^{\Lambda}\text{d}k k^{d-1}\int\text{d}\theta_1...\text{d}\theta_{d-2}\int_0^{\pi}\text{d}\theta\text{sin}^{d-2}\theta
\frac{\text{cos}^2\theta}{1+U\text{cos}^2\theta}\\
&=&\frac{S_{d-1}}{S_d}
\int_{0}^{\pi}\text{d}\theta\text{sin}^{d-2}\theta
\frac{\text{cos}^2\theta}{1+U\text{cos}^2\theta}
\end{eqnarray}
\end{mathletters}
where $S_d$ is the surface of the d-dimensional sphere $S_d=\frac{2\pi^{d/2}}{\Gamma(d/2)}$.
We finally find
\be
\label{Weff}
W^*(\Upsilon)=\langle w(\Upsilon)\rangle-\frac{1}{2}\frac{C\Upsilon^2}{A-B\Upsilon^2}\left[1-\frac{A}{A-B\Upsilon^2}
\frac{S_{d-1}}{S_d}I(d,U)\right]
\ee
where
\be
\label{integra}
I(d,U)=\int_0^{\pi}\text{d}\theta\text{sin}^{d-2}\theta\frac{\text{cos}^2\theta}{1+U\text{cos}^2\theta}
\ee
These expressions (\ref{Weff}) and (\ref{integra}) are our main results. They represent the expansion of the effective constant to second order for any type of non-linearities.\\
In the next section (\ref{SNL}), we apply this result to the case of SNL and we will recover the results of Bergman and Lee \cite{lee}. In section (\ref{WNL}), we apply our result to the WNL case.

\subsection{Application to the strongly non-linear case}
\label{SNL}

In the case of SNL, where $D=\chi(x)(E^2)^{\kappa/2}$ the energy density is given by $w(D)=a(x)(D^2)^{\nu/2}$ where $\nu=\frac{\kappa+2}{\kappa+1}$ and $a=(\chi)^{-\frac{1}{\kappa+1}}$. We recall here that in this method we have to express the energy density in terms of the divergence free field (i.e. $D$) and not $E$. The tensor $M$ here reads
\be
M_{ij}^{\alpha\gamma}=\delta_{\alpha\gamma}(A\delta_{ij}-B{\Upsilon}_i{\Upsilon}_j)-\beta C{\Upsilon}_i{\Upsilon}_j
\ee
where
\begin{mathletters}
\begin{eqnarray}
A&=&\nu\langle a\rangle{\Upsilon}^{\nu-2}\\
B&=&\nu(2-\nu)\langle a\rangle{\Upsilon}^{\nu-4}\\
C&=&\nu^2\langle\delta a^2\rangle{\Upsilon}^{2\nu-4}
\end{eqnarray}
\end{mathletters}
and $U=\frac{2-\nu}{\nu-1}=\kappa$. Here and in the following $\langle\delta XY\rangle=\langle XY\rangle-\langle X\rangle\langle Y\rangle$. We see that $M$ is positive definite, only if $\beta C$ is small enough, which is the condition for a low disorder expansion. We obtain (using (\ref{Weff})

\begin{eqnarray}
a^*&=&\frac{W^*}{(\Upsilon^2)^{\nu/2}}\\
&=&\langle a\rangle-\frac{\nu}{2(\nu-1)}\frac{\langle\delta a^2\rangle}{\langle a\rangle}
\left[
1-\frac{1}{\nu-1}\frac{S_{d-1}}{S_d}I(d,\kappa)
\right]
\end{eqnarray}
Given that $a^*=(\chi^*)^{-\frac{1}{\kappa+1}}$, we find the following perturbation expansion for $\chi^*$ for any space dimension (see appendix B for details)

\be
\label{chieff}
\chi^*=
\langle\chi\rangle-
\frac{1}{2}
\frac{\langle\delta\chi^2\rangle}{\langle\chi\rangle}
(\kappa+2)
\frac{S_{d-1}}{S_d}I(d,\kappa)+o(\langle\delta\chi^2\rangle)
\ee
This is the d-dimensional result in the SNL case. It is the same expression as  in \cite{lee} (this can be seen using the change of variables $u=\text{cos}\theta$, and $v^2=1-\frac{1-u^2}{1-\frac{\kappa}{\kappa+1}u^2}$. For $d=2$, the result (\ref{chieff}) reads
\be
\chi^*=
\langle\chi\rangle-
\frac{\langle\delta\chi^2\rangle}{\langle\chi\rangle}
\frac{\kappa+2}{2\kappa}
(1-\frac{1}{\sqrt{1+\kappa}})
\ee
and for $d=3$
\be
\chi^*=
\langle\chi\rangle-
\frac{\langle\delta\chi^2\rangle}{\langle\chi\rangle}
\frac{\kappa+2}{2\kappa}
(
1-\frac{1}{\sqrt{\kappa}}\text{arcsin}\sqrt{\frac{\kappa}{\kappa+1}}
)
\ee


\subsection{Application to the weakly non-linear case}
\label{WNL}

 In the WNL case, the energy density is $w=\varepsilon(x)(E^2)+\chi(x)(E^2)^2$ which in terms of $D$ gives $w(D)=a(x)(D^2)+b(x)(D^2)^2+{\cal O}(D^6)$ with $a(x)=1/\varepsilon(x)$ and $b(x)=-\chi(x)/\varepsilon^4$. We note that although $b(x)<0$, there are higher-order terms which are positive and guarantee the convergence of the integrals. We note that in this case we can go beyond perturbation theory but we will present these results elsewhere \cite{boprepa}.\\
In this case, we find
\begin{mathletters}
\begin{eqnarray}
A&=&\langle a\rangle-\langle b\rangle\Upsilon^2\\
B&=&2\langle b\rangle\\
C&=&\langle\delta(a-b\Upsilon^2)^2\rangle
\end{eqnarray}
\end{mathletters}
and $U=\frac{B\Upsilon^2}{A-B\Upsilon^2}$.\\
In general, the effective energy will not be a polynomial of the form $W^*=a^*\Upsilon^2+b^*\Upsilon^4$. In order to identify the effective coefficients, we have to expand the quantity $W^*$ to fourth order in $\Upsilon$. We obtain
\be
a^*=\langle a\rangle-\frac{\langle\delta a^2\rangle}{\langle a\rangle}(1-\frac{1}{d})
\ee
and
\be
b^*=\langle b\rangle-2(1-\frac{1}{d})\left[
(3-\frac{2}{d+2})\frac{\langle b\rangle}{\langle a\rangle}\frac{\langle\delta a^2\rangle}{\langle a\rangle}
-2\frac{\langle\delta ab\rangle}{\langle a\rangle})
\right]
\ee
We see in these expressions that for $d=1$, we obtain $a^*=\langle a\rangle$ and $b^*=\langle b\rangle$. This is the exact result since for $d=1$, the constraints $\nabla\cdot\sigma =0$ and $\overline{\sigma}=\Upsilon$ imply $\sigma =const=\Upsilon$.\\
Knowing that $a=1/\varepsilon$ and $b=\chi/\varepsilon^4$, we obtain after straightforward but tedious calculations
\be
\varepsilon^*\simeq\langle\varepsilon\rangle-\frac{1}{d}\frac{\langle\delta\varepsilon^2\rangle}{\langle\varepsilon\rangle}
\ee
and
\be
\chi^*\simeq\langle\chi\rangle\left[
1-4(2-\frac{1}{d})\frac{\langle\delta\varepsilon\chi\rangle}{\langle\varepsilon\rangle\langle\chi\rangle}+
2(\frac{10d^2+15d-16}{d(d+2)})
\frac{\langle\delta\varepsilon^2\rangle}{\langle\varepsilon\rangle}
\right]
\ee
These are the second-order perturbation results for the effective coefficients in the WNL case. We note the surprising result that, at this order, $\varepsilon^*$ does not depend on moments of $\chi$ and that $\chi^*$ does not depend on $\langle\delta\chi^2\rangle$. This suggests that the effective medium result for $\varepsilon^*$ should be independent of the non-linearity (in this weak non-linearity limit) and that $\chi^*$ is not determined self-consistently but is a function of $\varepsilon^*$ (this seems to be the case in a self-consistent resummation of the perturbation theory \cite{boprepa}).


\section{Conclusion}
\label{conclusion}
In this article, we proposed a new method to compute effective properties of disordered non-linear media for any type of non-linearities. In the case of strong non-linearity, we recover the known perturbation results and we present the second order perturbation result for the weakly non-linear case. We would like to emphasize that our method is general and could be used in many other situations (such as plasticity of porous media or the random fuse network). Moreover, this framework could be used as a starting point for more refined approximations.


\acknowledgements

One of us (MB) wants to thank D.J. Bergman for useful correspondence and Y.-P. Pellegrini for stimulating discussions.

\appendix
\section{}
In this appendix, we compute the path integral in expression (\ref{pathintper}). For this purpose, we will use the Fourier representation of the field $\epsilon_{\alpha}(x)$ (we keep the same notation for the field and its Fourier transform)
\be
\epsilon_{\alpha}(x)=\sum_k\tilde{\epsilon}_{\alpha}(k)e^{ik\cdot x}
\ee
The integral (\ref{pathintper}) can thus be rewritten as
\be
\int\text{d}^d\tilde{\epsilon}_{\alpha}(0)e^{-\beta\tilde{\epsilon}(0)\cdot M\cdot\tilde{\epsilon}(0)}
\int\prod_{k\neq 0}\text{d}^d\tilde{\epsilon}_{\alpha}(k)\delta(k\cdot\tilde{\epsilon}_{\alpha}(k))e^{-\beta\sum_{k\neq 0}\tilde{\epsilon}(k)\cdot M\cdot\epsilon(-k)}
\ee
By exponentiating the $\delta$-functions, one finds after integration the result (Eq. (\ref{perturb})) of the main text

\be
\langle Z^n\rangle\simeq
e^{\text{ln}\langle e^{-\beta n w(\Upsilon)}\rangle}
e^{-\frac{1}{2}\sum_{k\neq 0}\text{trln}M}
e^{-\frac{1}{2}\sum_{k\neq 0}\text{trln}(kM^{-1}k)}
\ee


\section{}
We explicit here the relation between the perturbation expansion of $a$ and that of $\chi$. We look for an expansion of $\chi^*$ of the form

\be
\label{eqchi}
\chi^*=\langle\chi\rangle-\frac{\langle\delta \chi^2\rangle}{\langle\chi\rangle}g_{\chi}
\ee
We expand this expression in power of the contrast $\Delta=\chi_2-\chi_1$, and find
\be
\label{eqchidev}
\chi^*\simeq\chi_1+q\Delta-\frac{pq\Delta^2}{\chi_1}g_{\chi}
\ee
Now, assuming that the perturbation expansion for $a$ is of the form
\be
\label{eqa}
a^*=\langle a\rangle-\frac{\langle\delta a^2\rangle}{\langle a\rangle}g_a
\ee
this leads to (to second order in $\Delta$)
\be
\label{eqa1}
a^*\simeq\chi_1^{-1/(\kappa+1)}
\left[
1-\frac{q\Delta}{\chi_1(\kappa+1)}+\frac{q\Delta^2}{\chi_1^2(\kappa+1)^2}
(\frac{\kappa+2}{2}-pg_a)
\right]
\ee
On the other hand, the relation between the functions $g_a$ is $g_{\chi}$ is found using $a^*=(\chi^*)^{-1/(\kappa+1)}$. Expanding this relation to second order in $\Delta$, we find
\be
\label{eqa2}
a^*\simeq \chi_1^{-1/(\kappa+1)}
\left[
1-\frac{q\Delta}{\chi_1(\kappa+1)}+\frac{q\Delta^2}{\chi_1^2(\kappa+1)}
(pg_{\chi}+\frac{q}{2}\frac{\kappa+2}{\kappa+1})
\right]
\ee
Comparing expressions (\ref{eqa1}) and (\ref{eqa2}), we obtain
\be
g_{\chi}=\frac{1}{2}\frac{\kappa+2}{\kappa+1}-\frac{g_a}{\kappa+1}
\ee
This relation allows to relate $g_{\chi}$ and $g_a$.



\end{document}